\documentclass[conference]{IEEEtran}
\IEEEoverridecommandlockouts

\usepackage{cite}
\usepackage{amsmath,amssymb,amsfonts}
\usepackage{algorithmic}
\usepackage{graphicx}
\usepackage{textcomp}
\usepackage{xcolor}
\def\BibTeX{{\rm B\kern-.05em{\sc i\kern-.025em b}\kern-.08em
    T\kern-.1667em\lower.7ex\hbox{E}\kern-.125emX}}

    \usepackage{ragged2e}
\usepackage{floatrow}
\usepackage{tabu}
\usepackage{multirow}
\usepackage{amsthm}
\usepackage[utf8]{inputenc}
\usepackage[english]{babel}
\theoremstyle{remark}
\usepackage[export]{adjustbox}

\usepackage{multicol}
\usepackage{epsfig}
\usepackage{epstopdf}
\usepackage{float}
\usepackage{perpage}
\MakeSorted{figure}
\MakeSorted{table}
\usepackage[normalem]{ulem}
\usepackage{array}
\usepackage{graphicx}
\usepackage{amsfonts}
\usepackage{amssymb}
\usepackage{soul}
\let\labelindent\relax
\usepackage{enumitem}							
\usepackage[english]{babel}
\usepackage[utf8]{inputenc}
\usepackage[linesnumbered,ruled,vlined]{algorithm2e}
\usepackage{upgreek}
\usepackage{bm} 
\usepackage{hyperref}
\usepackage{float}
\floatstyle{plaintop}
\restylefloat{table}
\hypersetup{linktocpage} 
\hypersetup{
colorlinks,
citecolor=black,
filecolor=black,
linkcolor=black,
urlcolor=black
}
\usepackage{tikz}
\usetikzlibrary{arrows.meta}
\usepackage{pifont}
\usepackage{makecell}


\usepackage{pgfplots}

\usepackage{cite}
\DeclareMathAlphabet\mathbfcal{OMS}{cmsy}{b}{n}
\ifCLASSINFOpdf
\else
\fi

\usepackage{array}
\usepackage{fixltx2e}

\usepackage[linesnumbered,ruled,vlined]{algorithm2e}

\usepackage[font=footnotesize]{caption} 
\usepackage{subcaption}
\usepackage{fix-cm}
\usepackage{comment}

\usepackage{mathtools}
\usepackage{pifont}

\usepackage{mathtools}
\usepackage{pifont}

\usepackage{tikz}

\begin{document}
\bstctlcite{IEEEexample:BSTcontrol}

\title{Joint Localization and Synchronization in\\Downlink Distributed MIMO}
\author{\\[-20pt]\IEEEauthorblockN{{Sauradeep Dey\IEEEauthorrefmark{1}, Musa Furkan Keskin\IEEEauthorrefmark{1}, Dario Tagliaferri\IEEEauthorrefmark{2}, Gonzalo Seco-Granados\IEEEauthorrefmark{3}, Henk Wymeersch\IEEEauthorrefmark{1}}}
\IEEEauthorblockA{
\IEEEauthorrefmark{1}Department of Electrical Engineering, Chalmers University of Technology, Sweden\\
\IEEEauthorrefmark{2}Department of Electronics, Information and Bioengineering, Politecnico di Milano, Italy\\
\IEEEauthorrefmark{3}Universitat Autonoma de Barcelona, Spain
}
\thanks{This work was partly supported by the SNS JU project 6G-DISAC under the EU’s Horizon Europe research and innovation programme under Grant Agreement No 101139130, the Swedish Research Council (VR) through the project 6G-PERCEF under Grant 2024-04390, the Spanish R+D project PID2023-152820OB-I00, and the AGAUR-ICREA Academia Program.} 
\vspace{-0.8cm}
}

\maketitle

\begin{abstract}
We investigate joint localization and synchronization in the downlink of a distributed multiple-input-multiple-output (D-MIMO) system, aiming to estimate the position and phase offset of a single-antenna user equipment (UE) using downlink transmissions of multiple phase-synchronized, multi-antenna access points (APs). We propose two transmission protocols: sequential (\textit{P1}) and simultaneous (\textit{P2}) AP transmissions, together with the ML estimators that either leverage  (coherent estimator) or disregard phase information (non-coherent estimator). Simulation results reveal that downlink D-MIMO holds significant potential for high-accuracy localization while showing that \textit{P2} provides superior localization performance and reduced transmission latency. 
\end{abstract}

\vspace{-3mm}
\section{Introduction}
Localization has been a key research area in wireless communications, evolving from early cellular systems to advanced techniques in 5G and beyond \cite{gonzaloSurvey_2018}. Traditional methods such as time difference of arrival (TDoA) and angle of arrival (AoA) have been widely used, but their accuracy was limited by narrow bandwidth, hardware timing offsets and small arrays at base stations (BSs)  \cite{mogyorosi2022positioning}. The advent of massive MIMO in 5G introduced large antenna arrays, improving localization accuracy by leveraging high spatial resolution to estimate AoA and angle-of-departure (AoD) \cite{mMIMO_2019}. Looking ahead to 6G, localization is envisioned as a core service with stringent requirements (e.g., sub-cm level for immersive telepresence \cite{6g_hexax}). To fulfill such demands in 6G, distributed MIMO (D-MIMO), where multiple spatially distributed access points (APs) cooperate to function as a unified system, has emerged as a promising approach for localization \cite{guo2024integrated}. By exploiting near-field propagation effects induced by phase coherence among APs, D-MIMO systems can yield extreme accuracies that can support 6G use cases such as digital twinning and extended reality \cite{VTM_gaps_6G_2023}.

\begin{figure}
    \centering
    \includegraphics[width=0.99\linewidth]{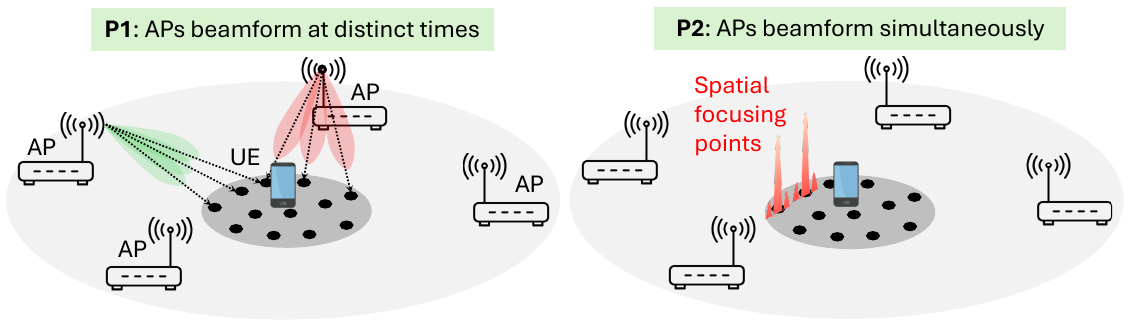}
    \vspace{-5mm}
    \caption{System model of a downlink D-MIMO system for UE localization. The APs are phase-synchronized, leading to two options for downlink beamforming (\textit{P1} and P2).}
    \label{fig:SystemModel}
\end{figure}

A distinguishing feature of emerging D-MIMO localization is the possibility of phase-coherent operation across distributed APs, enabling the entire D-MIMO system to form an extremely large aperture array, which introduces wavefront curvature (i.e., near-field) effects and tremendous gains in resolution \cite{vukmirovic2018position}. Coherent processing in D-MIMO systems can push the position error to orders of magnitude below the carrier wavelength \cite{vukmirovic2021performance}. Recent studies have mainly focused on \textit{uplink}-based D-MIMO localization \cite{vukmirovic2018position,vukmirovic2019direct,vukmirovic2021performance,RS_UPLINK_JOURNAL}, where the network-side phase-coherent processing uses uplink signals received at multiple APs to infer the location of a transmitter user equipment (UE). 


Extending this framework to downlink offers  advantages and challenges. The key benefit lies in beamfocusing \cite{demir2021foundations}, where multiple APs can coordinate to concentrate signals around the UE, unlike uplink, where a UE can only beamform toward a single AP. Downlink also scales efficiently for multiple UEs without additional network overhead, while uplink requires dedicated resources per UE. Additionally, downlink enables decentralized localization, allowing each UE to compute its position locally, reducing fronthaul data load \cite{guo2024integrated}. The primary challenge is that downlink places the processing burden on the UE, which has limited capabilities and must process superposed signals from multiple APs. Despite these trade-offs, phase-coherent downlink D-MIMO localization remains unexplored, leaving a  gap in the field.

In this  paper, we make progress towards filling this gap by addressing the fundamental problem of joint localization and synchronization of a UE using downlink signals transmitted by spatially distributed phase-coherent APs. The main contributions are as follows: 
\textit{(i)} For the first time, we propose a downlink phase-coherent D-MIMO localization system, 
using signals transmitted by multiple phase-synchronized multi-antenna APs. For the transmit signal design, we propose two different transmission protocols: \textit{P1}, where the APs transmit their signals sequentially for conventional far-field beamforming, and \textit{P2}, where all the APs transmit simultaneously for near-field beamfocusing.
    \textit{(ii)} For both transmission protocols, we design the maximum likelihood (ML) estimators for position and phase offset estimation, considering both coherent and non-coherent processing approaches. A hybrid two-step optimization strategy is developed that capitalizes on non-coherent processing to provide favorable initial estimates for high-resolution coherent processing. 
    \textit{(iii)} We carry out a  performance evaluation of the designed estimators alongside the corresponding performance bounds, showcasing the potential of phase-coherent D-MIMO architectures for high-accuracy localization. Additionally, we observe that \textit{P2} outperforms \textit{P1} in both accuracy and latency.

\vspace{-3mm}
\section{System model} \label{sec:system_model}
{We consider downlink of a D-MIMO system, as illustrated in Fig.~\ref{fig:SystemModel}, where $M$ randomly deployed APs transmit narrowband signals, to enable the localization and synchronization of a user equipment (UE). The  system operates at frequency $f_0$ with (small) bandwidth $W$.\footnote{In particular, $c/W \gg \max_m \Vert \mathbf{p}_m-\mathbf{u}\Vert-\min_m \Vert \mathbf{p}_m-\mathbf{u}\Vert$ so that no position information can be extracted from signal bandwidth. The extension to multi-carrier (wideband) systems is not treated here. The extension to multiple UEs is also not considered, as it is trivial.}  The 2D location of the $m$-th AP is given as $\mathbf{p}_m=[x_m,y_m]^\top$, while the location of the UE is given as $\mathbf{u}=[u_x,u_y]^\top$. Each AP is equipped with a uniform linear array (ULA) of $N$ antennas, spaced by $\lambda_0/2$, where $\lambda_0=c/f_0$ is the carrier wavelength and $c$ is the speed of light, while the UE has a single antenna. The orientation of AP $m$ is denoted by $o_m\in [0,2\pi)$, measured counterclockwise with respect to the x-axis.
The APs are phase-synchronized, thus behaving like a distributed array~\cite{RS_UPLINK_JOURNAL}, while the UE is assumed to have an unknown phase offset $\delta_\phi$, with respect to the AP network. The goal of the system is to estimate both the UE position $\mathbf{u}$ and the phase offset $\delta_\phi$.  
Without loss of generality, the UE is assumed to be in the far-field of each individual AP, but in the \textit{near-field} of the complete AP network.}

\subsection{{{Channel Model}}}\label{subsec:channel_model} 
{The channel between the $m$-th AP and the UE is represented by the vector $\mathbf{h}_{m}\in \mathbb{C}^{N\times 1}$, which is assumed herein to have a dominant LoS component, and is given as}
\begin{align}
    \mathbf{h}_{m} = \alpha_m e^{j \delta_\phi} e^{-j2\pi f_0\tau_m}\mathbf{a}({\theta_m})
    \triangleq    \alpha_me^{j{\phi_m}}\mathbf{a}({\theta_m}).\label{eq:channel_1}
\end{align}
Here,  $\alpha_m \in \mathbb{R}$ denotes the large scale fading coefficient, modeled as $\alpha_m = \lambda_0/(4 \pi \| \mathbf{u}-\mathbf{p}_m \|$),  $\tau_m = \|\mathbf{u} - \mathbf{p}_m \|/c$ is the signal time of arrival (TOA) from the $m$-th AP to the UE, $\mathbf{a}({\theta_m})\in\mathbb{C}^{N\times 1}$ represents the $m$-th AP array response vector, with angle of departure (AOD) $\theta_m= \arctan((y_m-u_y)/(x_m-u_x))-o_m$.
The $n$-th element of the array response vector is $ [\mathbf{a}(\theta_m)]_n = \exp(j \pi n \sin \theta_m)$. 
In \eqref{eq:channel_1},
$\phi_m \triangleq -2\pi f_0 \tau_m + \delta_{\phi}$ represents the carrier phase term, that accounts for the effects of signal propagation. It depends on the UE position $\mathbf{u}$ through $\tau_m$, and the phase offset~$\delta_\phi$.

\subsection{{{Signal Transmission Protocols}}}\label{subsec:signal_tx_protocol} 
{The UE localization and synchronization problem considered here exploits multiple transmissions over time. 
The $m$-th AP transmits the symbol $s_m[k]$ at the $k$-th time instant, $k=0,...,K-1$, using the precoder $\mathbf{f}_m[k]\in\mathbb{C}^{N\times 1}$, where the transmitted symbol and the precoder satisfies the power constraint $\mathbb{E}\{|s_m[k]|^2\} =
\mathbb{E}\{\|\mathbf{f}_m[k]\|^2\}=1$. In this work, we analyze two different transmission protocols (see Fig.~\ref{fig:SystemModel}), referred to as \textit{P1} and \textit{P2}, which dictate how APs coordinate their operation. These transmission protocols are discussed in detail below.} For both protocols, we consider $K$ illumination points, based on the prior knowledge of the UE location. We denote the $k$-th 2D illumination point by $\mathbf{i}[k$].

\subsubsection{Transmission Protocol \textit{P1}}
In this transmission protocol, the APs transmit their signals using time-division multiplexing (TDM), ensuring that at any given time instant, the UE receives the signal from only one AP, which requires $K$ transmissions per AP. 
Hence, the total transmission duration is $T_{\rm P1} = MKT$, where $T=1/W$ represents the duration of a single transmission (or symbol). The 
 signal received by the UE at time $k$ from the $m$-th AP is given by
\begin{align} \nonumber
    y_m[k] &= \sqrt{P_m} \, \mathbf{h}_m^{\top} \mathbf{f}_m[k] s_m[k] + n_m[k] \,, \\[2pt]
    & = \alpha_m e^{j\delta_{\phi}} x_m[k] + n_m[k]    \label{eq:rx_sig_single_sub_P1}
\end{align} 
for $k \in \mathcal{T}_m, m = 1, \ldots, M$, where $P_m$ is the power  transmitted by the $m$-th AP and $\mathbf{f}_m[k]=e^{j 2 \pi f_0 \tau_m[k]} \mathbf{a}^*({\theta_m[k]})/\sqrt{N}$, where 
$\theta_m[k]= \arctan((y_m-i_y[k])/(x_m-i_x[k]))-o_m$ and $\tau_m[k]=\|\mathbf{i}[k] - \mathbf{p}_m \|/c$. The set $\mathcal{T}_m = \{m, m+M, m+2M,...,m+(K-1)M\}$ denotes the set of time instants when the $m$-th AP transmits its signal. The scalar $n_m[k]\sim \mathcal{CN}(0,\sigma_n^2)$ is the additive white Gaussian noise (AWGN) at the $k$-th time instant within $\mathcal{T}_m$ and $x_m[k]$  represents the transmit signal after propagation through the channel, excluding the channel amplitude and the phase offset and
 is defined as 
\begin{align}\label{eq:x_m_k}
x_m[k]=\sqrt{P_m}e^{-j2\pi f_0\tau_m} \mathbf{a}^{\top}({\theta_m})\mathbf{f}_m[k] s_m[k].\nonumber    
\end{align}

\subsubsection{Transmission Protocol \textit{P2}}
In protocol \textit{P2}, the AP network \textit{coherently focuses} the signal on a desired spot within the target region. Over $K$ transmissions, the illumination point is changed to cover the entire area of interest, resulting in a total transmission duration of $T_{\rm P2} = KT$. This is in contrast to protocol \textit{P1}, where the transmission time scales with the number of APs $M$, resulting in $T_{\rm P2} \ll T_{\rm P1}$, for a fixed $K$.
The downlink signal received by the UE at the $k$-th time instant~is 
\begin{align} \nonumber
    y[k]
    &=\sum_{m=1}^{M}\sqrt{P_m} \, \mathbf{h}_m^{\top} \mathbf{f}_m[k] s_m[k] + n[k] \,,\\
    &= \sum_{m=1}^{M}\alpha_m e^{j\delta_{\phi}}{x}_m[k]+n[k] \,,
\end{align}
where the precoders $\mathbf{f}_m[k]$ are identical to those in \textit{P1}.



\section{Joint Localization and Synchronization}\label{sec:parameter_estimation}
In this section, we estimate the 2D UE position $\mathbf{u}$ and the phase offset $\delta_\phi$ for both the transmission protocols. The UE receives the signal $\mathbf{y} = [y[0], \dots, y[S-1]]^\top \in \mathbb{C}^{S \times 1}$ over $S$ transmissions, according to the respective transmission protocols (i.e., $S=MK$ for \textit{P1} and $S=K$ for \textit{P2}). 
Given the parameter vector to be estimated, $\bm{\eta} = [\mathbf{u}^\top, \delta_\phi, \bm{\alpha}^\top]^\top \in \mathbb{R}^{(M+3) \times 1}$, the joint localization and synchronization of the UE is formulated as  an ML estimation problem
\begin{align}\label{eq: ML_opt1}
   \widehat{\bm{\eta}}
   &=\underset{\bm{\eta}}{\text{ argmax }} \log p(\mathbf{y}|\bm{\eta}),
\end{align}
{where $p(\mathbf{y}|\bm{\eta})$ represents the likelihood function. 
Notice that estimating the parameters of interest, $\mathbf{u}$ and $ \delta_\phi $, also requires estimating the propagation amplitudes $\bm{\alpha} = [\alpha_1, \cdots, \alpha_M]^\top \in \mathbb{R}^{M \times 1}$, which act as nuisance parameters. For each transmission protocol, we design two estimators: \textit{(i)} a \textit{coherent} estimator that exploits the carrier phase information; and \textit{(ii)} a \textit{non-coherent} estimator that does not use the carrier phase.  
\vspace{-2mm}
\subsection{{Transmission Protocol \textit{P1}}}
\subsubsection{Coherent ML Estimation}\label{subsubsec_3xp_CP_inf_P1}
{We formulate the UE localization and synchronization problem as the joint estimation problem which uses the carrier phase, i.e., the explicit dependence of the carrier phase $\phi_m$ on the UE position $\mathbf{u}$ through the TOA $\tau_m$ and phase offset $\delta_\phi$. 
The problem \eqref{eq: ML_opt1} becomes
\begin{align}\label{eq: ML_opt_P1_coherent}
   \widehat{\bm{\eta}}
   &= \underset{\bm{\eta}}{\text{ arg min }} \sum\limits_{m=1}^{M}\sum\limits_{k\in \mathcal{T}_m}|y[k]-e^{j\delta_{\phi}}\alpha_m x_m[k]|^2.
\end{align}
We tackle this optimization problem by first solving for the optimal values of $\alpha_m$ for $m = 1, \dots, M$. Differentiating the cost function in \eqref{eq: ML_opt_P1_coherent} with respect to $\alpha_m$ and equating it to zero, we get  
\begin{align}\label{eq:alpha_opt_P1_coherent}
    \hat{\alpha}_m = \frac{\sum\limits_{k\in\mathcal{T}_m}\mathcal{R}\big\{e^{j\delta_\phi}x_m[k] y^*[k]\big\}}{\sum\limits_{k\in\mathcal{T}_m}|x_m[k]|^2}
    \triangleq \frac{\mathcal{R}\big\{e^{j\delta_\phi}\mathbf{y}_m^{\mathsf{H}}\mathbf{x}_m\big\}}{\|\mathbf{x}_m\|^2_2} \,,
\end{align}
{where, the vectors $\mathbf{y}_m=[y[k]]_{k\in\mathcal{T}_m}$ and $\mathbf{x}_m=[x_m[k]]_{k\in\mathcal{T}_m}$ gathers the samples of the observation and signal model for the $m$-th AP, respectively. Note that $\hat{\alpha}_m$ is \textit{conditional} on $\mathbf{u}$ and $ \delta_\phi $.
Substituting $\hat{\alpha}_m$   into \eqref{eq: ML_opt_P1_coherent}, we obtain}
\begin{align}
    &\sum\limits_{m=1}^{M}\sum\limits_{k\in \mathcal{T}_m}\Big|y[k]-e^{j\delta_{\phi}}\frac{\mathcal{R}\big\{e^{j\delta_\phi}\mathbf{y}_m^{\mathsf{H}}\mathbf{x}_m\big\}}{\|\mathbf{x}_m\|^2} x_m[k]\Big|^2\nonumber\\
    = &\sum\limits_{m=1}^{M}\sum\limits_{k\in \mathcal{T}_m}\Big|\underset{\widetilde{y}_m[k]}{\underbrace{y[k]-\frac{(\mathbf{x}_m^{\mathsf{H}} \mathbf{y}_m)x_m[k]}{2\|\mathbf{x}_m\|^2}}}-e^{j2\delta_{\phi}}\underset{c_m[k]}{\underbrace{\frac{(\mathbf{y}_m^{\mathsf{H}}\mathbf{x}_m)x_m[k]}{2\|\mathbf{x}_m\|^2}}} \Big|^2\nonumber\\[-10pt]
    = &\sum\limits_{m=1}^{M}\|\widetilde{\mathbf{y}}_m-e^{j2\delta_{\phi}}\mathbf{c}_m \|^2,\label{eq: ML_obj_P1_coherent_2}
\end{align}
{where $\widetilde{\mathbf{y}}_m=[\widetilde{y}_m[k]]_{k\in\mathcal{T}_m}$, and $\mathbf{c}_m=[c_m[k]]_{k\in\mathcal{T}_m}$.
The value of $\delta_\phi$ that maximizes the objective in \eqref{eq: ML_obj_P1_coherent_2} is given as} $ \hat{\delta}_\phi=\frac{1}{2}\angle{\sum_{m=1}^M\mathbf{c}_m^{\mathsf{H}}\widetilde{\mathbf{y}}_m}$, 
which remains conditional on $\mathbf{u}$.
{By substituting $\hat{\delta}_\phi$ in \eqref{eq: ML_obj_P1_coherent_2}, the ML optimization problem becomes}
\begin{align}\label{eq:ML_opt_final_P1_coherent}
    \hat{\mathbf{u}}=\underset{\mathbf{u}}{\text{ arg min }} \sum\limits_{m=1}^{M}\|\widetilde{\mathbf{y}}_m(\mathbf{u})-e^{j2\hat{\delta}_{\phi}(\mathbf{u})}\mathbf{c}_m(\mathbf{u}) \|^2.
\end{align}
The optimal UE position cannot be obtained in closed form. A practical approach to solve \eqref{eq:ML_opt_final_P1_coherent} is to perform a  2D grid search.


\subsubsection{Non-coherent ML Estimation}\label{subsubsec_wo_3xp_CP_inf_P1}
An alternative approach to solve the joint localization and synchronization problem is the non-coherent one, where the carrier phase $\phi_m$ is assumed to be an unknown (nuisance) parameter that does not depend on the UE position $\mathbf{u}$.
In this case the channel amplitudes are treated as unstructured complex variables $\beta_m=\alpha_m e^{j\phi_m}$. The corresponding ML optimization problem is formulated as 
\begin{align}\label{eq: ML_opt_P2_coherent1}
[\hat{\mathbf{u}},\hat{\bm{\beta}}]   = \underset{\{\mathbf{u},\bm{\beta}\}}{\text{ arg min }} \sum\limits_{m=1}^{M}\sum\limits_{k\in \mathcal{T}_m}\left|y[k]-\beta_m z_m[k]\right|^2.
\end{align}
where $z_m[k]=\sqrt{P_m}\mathbf{a}^{\top}({\theta_m})\mathbf{f}_m[k] s_m[k]$ is the signal model excluding the complex amplitudes. The  estimate of $\beta_m$ conditional on $\mathbf{u}$ is given by $\hat{\beta}_m=\mathbf{z}_m^{\mathsf{H}}\mathbf{y}_m/\|\mathbf{z}_m\|^2$, 
where $\mathbf{z}_m=[z_m[k]]_{k\in\mathcal{T}_m}$ gathers the samples of the signal model for the $m$-th AP. By substituting $\hat{\beta}_m$ in \eqref{eq: ML_opt_P2_coherent1}, we obtain the non-coherent ML estimator}
\begin{align}
\label{eq: ML_opt_P2_coherent2}
\hat{\mathbf{u}}   = \underset{\mathbf{u}}{\text{ argmin }} \sum\limits_{m=1}^{M}\Big\|\mathbf{y}_m-\frac{\mathbf{z}_m(\mathbf{u})^{\mathsf{H}}\mathbf{y}_m}{\|\mathbf{z}_m(\mathbf{u})\|^2} \mathbf{z}_m(\mathbf{u})\Big\|^2,
\end{align}
which can be solved by a  2D grid search. 



\vspace{-3mm}
\subsection{{Transmission Protocol P2}}
\subsubsection{Coherent ML Estimation}
Similar to Section \ref{subsubsec_3xp_CP_inf_P1}, we design the coherent ML estimator for UE localization and synchronization, by exploiting the carrier phase for transmission protocol \textit{P2}. By gathering the signal models $x_m[k]$ $\forall\,k,m$ from \eqref{eq:x_m_k} into matrix $\mathbf{X}\in\mathbb{C}^{K\times M}$ and defining $\mathbf{y}=[y[1],\cdots,y[K]]^\top$ and $\bm{\alpha}=[\alpha_1,\cdots,\alpha_M]^\top$, the ML estimation problem is 
\begin{align} \nonumber
   \widehat{\bm{\eta}}
    &= \underset{\bm{\eta}}{\text{ arg min }} \|\mathbf{y}-e^{j\delta_{\phi}}{\mathbf{X}}\bm{\alpha}\|_2^2 \,,\\
   &=\underset{\bm{\eta}}{\text{ arg min }} -2\mathcal{R}\{e^{-j\delta_{\phi}}\bm{\alpha}^\top{\mathbf{X}}^{\mathsf{H}}\mathbf{y}\}+\bm{\alpha}^\top{\mathbf{X}}^{\mathsf{H}}{\mathbf{X}}\bm{\alpha} \,.
   \label{eq: ML_opt_coherent_P2}
\end{align}
By applying {\cite[Lemma S1]{MONOSTATIC_FURKAN}}, the second term in the objective of \eqref{eq: ML_opt_coherent_P2} simplifies to $\bm{\alpha}^\top\mathcal{R}\{{\mathbf{X}}^{\mathsf{H}}{\mathbf{X}}\}\bm{\alpha}$. Substituting this and using the fact that $\bm{\alpha}$ is a real vector, we rewrite \eqref{eq: ML_opt_coherent_P2} as 
\begin{align} 
   \widehat{\bm{\eta}}
   &=\underset{\bm{\eta}}{\text{ arg min }} \bm{\alpha}^\top\mathcal{R}\{{\mathbf{X}}^{\mathsf{H}}{\mathbf{X}}\}\bm{\alpha}-2\mathcal{R}\{e^{-j\delta_{\phi}}\bm{\alpha}^\top{\mathbf{X}}^{\mathsf{H}}\mathbf{y}\}.
   \label{eq: ML_opt_coherent_P2_3}
\end{align}
To obtain the optimal $\bm{\alpha}$, we differentiate the objective function with respect to $\bm{\alpha}$ and set it to zero. For a given $\delta_\phi$ value, the optimal $\bm{\alpha}$ is 
$\hat{\bm{\alpha}}
   =\mathcal{R}\{{\mathbf{X}}^{\mathsf{H}}{\mathbf{X}}\}^{-1}\mathcal{R}\{e^{-j\delta_{\phi}}{\mathbf{X}}^{\mathsf{H}}\mathbf{y}\}$.
{Substituting $\hat{\bm{\alpha}}$ in the objective function in \eqref{eq: ML_opt_coherent_P2}, we get}
\begin{align}
[\hat{\mathbf{u}},\hat{\delta}_{\phi}]
&=\underset{\{\mathbf{u},\delta_{\phi}\}}{\text{ arg min }}\|\mathbf{v}-e^{j2\delta_{\phi}}\mathbf{w}\|^2,\label{eq: ML_opt_coherent_P2_4}
\end{align}
{where, $\mathbf{v}= \mathbf{y}-\mathbf{X}\mathcal{R}\{\mathbf{X}^H\mathbf{X}\}^{-1}\mathbf{X}^H\mathbf{y}/2$, and $\mathbf{w}=\mathbf{X}\mathcal{R}\{\mathbf{X}^H\mathbf{X}\}^{-1}\mathbf{X}^T\mathbf{y}^*/2$. 
The optimal $\delta_\phi$ is obtained in closed form, and is $\hat{\delta}_\phi= \frac{\angle\mathbf{w}^H\mathbf{v}}{2}$. Substituting $\hat{\delta}_\phi$ in \eqref{eq: ML_opt_coherent_P2_4}, we get}
\begin{align}\label{eq: ML_opt_coherent_P2_5}
    \hat{\mathbf{u}}=\underset{\mathbf{u}}{\text{ arg min }}\|\mathbf{v}-e^{j2\hat{\delta}_{\phi}}\mathbf{w}\|^2
\end{align}
{To solve \eqref{eq: ML_opt_coherent_P2_5}, a 2D exhaustive search is performed.}
\begin{figure}
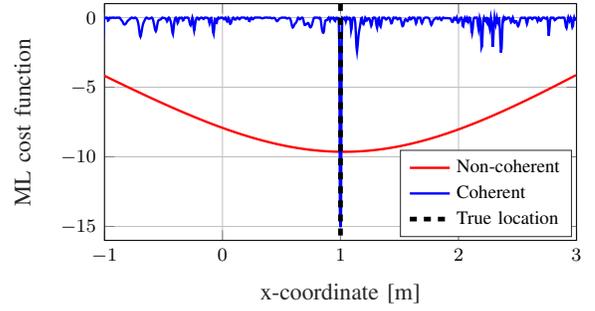

    \centering
    \include{TikzFigures/P1}
    \vspace{-3em}
        \caption{ML cost functions with respect to the x-coordinate of UE position in non-coherent and coherent processing for transmission protocol \textit{P1}.}
    \label{fig:P1-HW}    
\end{figure}

\subsubsection{Non-coherent ML Estimation}
{For non-coherent ML estimation, similar to Section \ref{subsubsec_wo_3xp_CP_inf_P1}, the carrier phase $\phi_m$ is treated as an unknown parameter independent of the UE position, and the channel amplitudes are considered as unstructured complex scalar quantities $\beta_m = \alpha_m e^{j\phi_m}$. The corresponding ML optimization problem is $\min_{\mathbf{u},\bm{\beta}} \|\mathbf{y}-{\mathbf{Z}}(\mathbf{u})\bm{\beta}\|^2$}
{where $\bm{\beta}=[\beta_1,\cdots,\beta_M]^\top\in\mathbb{C}^{M\times 1}$, and matrix $\mathbf{Z}(\mathbf{u})\in \mathbb{C}^{K \times M}$ gathers the model quantities $z_m[k]$. The optimal $\bm{\beta}$ can be readily derived as} $ \hat{\bm{\beta}}= \left({\mathbf{Z}}^{\mathsf{H}}{\mathbf{Z}}\right)^{-1}{\mathbf{Z}}^{\mathsf{H}}\mathbf{y}$.
{Substituting the optimal $\bm{\beta}$ in the ML cost function in \eqref{eq: ML_opt_non_coherent_P2}, we obtain}
\begin{align}
   \hat{\mathbf{u}}
   & = { \arg \min_{\mathbf{u}} } \| \mathbf{y} - \mathbf{Z}\left({\mathbf{Z}}^{\mathsf{H}}{\mathbf{Z}}\right)^{-1}{\mathbf{Z}}^{\mathsf{H}}\mathbf{y}\|^2, 
   \label{eq: ML_opt_non_coherent_P2}
\end{align}
which can be solved by a 2D grid search. 

\subsection{Coherent vs. Non-Coherent ML Cost-Functions}
In this section, we first analyze the differences between the coherent and non-coherent ML cost functions for  transmission protocol \textit{P1} (conclusions are similar for  P2). Based on this analysis, we develop a {low-complexity} algorithm to estimate the UE position,  combining the two cost functions. 

\subsubsection{ML Cost Function Analysis}

Fig.~\ref{fig:P1-HW} shows a  1D cut of  coherent and non-coherent ML cost functions for protocol \textit{P1}.  As expected, leveraging the carrier phase yields a cost function with a more pronounced global minimum at the true UE location, as compared to its non-coherent counterpart. However, the coherent ML cost function features several local minima that prevents the use of conventional gradient descent techniques, whereas the non-coherent ML cost function is smoother, with a distinct global minimum but yield less accurate UE position estimates.



\subsubsection{Proposed Hybrid Approach}
{To harness the strengths of both cost functions,
we propose a two-stage localization algorithm. In the first stage, we search for the minimum of the non-coherent cost function. Since this cost-function is smooth, efficient gradient descent with a coarse initial guess is sufficient.  This provides a good initial estimate, close to the  UE position. In the second stage, we refine the estimate by searching for the minimum of the coherent ML cost function. We perform this search only within a small region around the initial estimate, using a finer grid.}

\section{Simulation Results}\label{sec: Simulation_results}\vspace{0pt}
In this section, we perform numerical simulations to evaluate performances under both transmission protocols.

\subsection{Scenario}
We consider a network of $M = 20$ APs, each equipped with $N = 4$ antennas. The APs are placed uniformly along the edges of a $40$ m $\times$ $40$ m square centered at the origin, as shown in Fig.~\ref{fig:System_deployment}, with their antennas oriented towards the center. The APs precoded unity pilots (i.e., $s_m[k] = 1 ~ \forall m,k$) to illuminate $K = 36$ illumination points. These illumination points are uniformly distributed over a $5$ m $\times$ $5$ m square region around the origin. The total transmit energy of the AP network remains fixed across both transmission protocols.
The UE is positioned at $\mathbf{u} = [1,1]^\top$ m and has a phase offset of $\delta_\phi = 10^\circ$ relative to the AP network. The system operates at a carrier frequency of $f_0 = 3.5$ GHz. The total transmit energy of the network is fixed for both transmission protocols. The noise power is evaluated over a reference bandwidth of $W = 120$ kHz, corresponding to a single OFDM subcarrier spacing, and is given by  
$\sigma_n^2 = k_B T_{\rm{tn}} W,$ where $k_B = 1.38 \times 10^{-23}$ J/K is the Boltzmann constant, and $T_{\rm{tn}} = 290$ K is the standard thermal noise temperature.

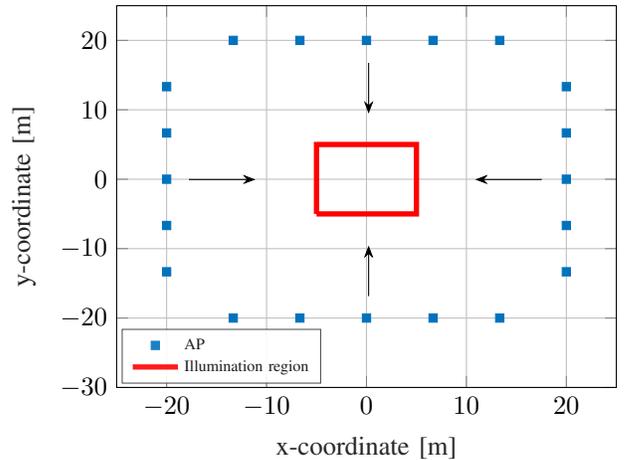
\begin{figure}[t]
\vspace{-200pt}
    \centering
%
%
\definecolor{mycolor1}{rgb}{0.00000,0.44700,0.74100}%
\begin{tikzpicture}

\begin{axis}[%
width=0.75\linewidth,
height=2in,
at={(0.737in,0.584in)},
scale only axis,
xmin=-25,
xmax=25,
xlabel style={font=\color{white!15!black}},
xlabel={x-coordinate [m]},
ymin=-30,
ymax=25,
ylabel style={font=\color{white!15!black}},
ylabel={y-coordinate [m]},
axis background/.style={fill=white},
xmajorgrids,
ymajorgrids,
legend style={font=\footnotesize,at={(0.01,0.01)},nodes={scale=0.75, transform shape}, anchor=south west, legend cell align=left, align=left, draw=white!25!black, opacity=0.9}
]
\addplot[only marks, mark=square*, mark options={}, mark size=1.5607pt, color=mycolor1, fill=mycolor1] table[row sep=crcr]{%
x	y\\
-13.3333333333333	-20\\
-6.66666666666667	-20\\
0	-20\\
6.66666666666667	-20\\
13.3333333333333	-20\\
20	-13.3333333333333\\
20	-6.66666666666667\\
20	0\\
20	6.66666666666667\\
20	13.3333333333333\\
13.3333333333333	20\\
6.66666666666667	20\\
0	20\\
-6.66666666666667	20\\
-13.3333333333333	20\\
-20	13.3333333333333\\
-20	6.66666666666667\\
-20	0\\
-20	-6.66666666666667\\
-20	-13.3333333333333\\
};
\addlegendentry{AP}

\addplot [color=red, line width=2.0pt]
  table[row sep=crcr]{%
-5	-5\\
-3	-5\\
-1	-5\\
1	-5\\
3	-5\\
5	-5\\
5	-3\\
5	-1\\
5	1\\
5	3\\
5	5\\
3	5\\
1	5\\
-1	5\\
-3	5\\
-5	5\\
-5	3\\
-5	1\\
-5	-1\\
-5	-3\\
-5	-5\\
};
\addlegendentry{Illumination region}

\end{axis}

\begin{axis}[%
width=6.969in,
height=5.312in,
at={(0in,0in)},
scale only axis,
xmin=0,
xmax=1,
ymin=0,
ymax=1,
axis line style={draw=none},
ticks=none,
axis x line*=bottom,
axis y line*=left
]
\draw[-{Stealth}, color=black] (axis cs:0.16,0.315) -- (axis cs:0.21,0.315);
\draw[-{Stealth}, color=black] (axis cs:0.295,0.20) -- (axis cs:0.295,0.25);
\draw[-{Stealth}, color=black] (axis cs:0.425,0.315) -- (axis cs:0.375, 0.315);
\draw[-{Stealth}, color=black] (axis cs:0.295,0.43) -- (axis cs:0.295,0.38);
\end{axis}
\end{tikzpicture}%
    \vspace{-4.5em}
        \caption{D-MIMO system deployment. The orientation of the AP antenna arrays are indicated by the arrows.}
    \label{fig:System_deployment}    
\end{figure}



\subsection{Algorithm Performance Assessment}\vspace{-5pt}
We begin by evaluating the performance of the designed estimators for transmission protocols \textit{P1} and \textit{P2}. Fig.~\ref{fig:RMSE_PEB_vs_tx_power} presents the root mean squared error (RMSE) of the UE position estimate $\mathbf{u}$ for both protocols, along with their corresponding PEBs, as a function of the per-AP transmit power. We see that both \textit{P1} and \textit{P2} achieve sub-meter accuracy even at low transmit power levels, with centimeter-level (or better) precision attained as the SNR increases.
We further observe that protocol \textit{P2}, where APs transmit simultaneously—resulting in lower latency—outperforms \textit{P1} in both RMSE and PEB. Specifically, the PEB with \textit{P2} is approximately $6$ dB lower than that of \textit{P1}. However, this improvement depends on the number of illumination points, which is set to $K = 36$ in this case. A more detailed analysis of the PEB as a function of the number of illumination points is provided later in Fig.~\ref{fig:PEB_vs_IP}. Additionally, as expected, the PEB for non-coherent processing is significantly higher than that for coherent processing—by approximately $24$ dB for \textit{P1} and $30$ dB for \textit{P2}. Furthermore, for non-coherent processing, the PEBs for both \textit{P1} and \textit{P2} are nearly identical for $K = 36$; however, for smaller $K$ values, \textit{P2} performs substantially worse than \textit{P1}, as will be investigated next.

\begin{figure}
\vspace{-0pt}
    \centering
%
%
\begin{tikzpicture}[scale=1\columnwidth/9cm,font=\footnotesize]
\begin{axis}[%
width=0.75\columnwidth,
height=1.8in,
at={(0.905in,0.583in)},
scale only axis,
xmin=-10,
xmax=40,
xlabel style={font=\color{white!15!black}},
xlabel={Per AP transmit power in dBm},
ymode=log,
ymin=1e-06,
ymax=2,
yminorticks=true,
ylabel style={font=\color{white!15!black}},
ylabel={RMSE  [m]},
axis background/.style={fill=white},
xmajorgrids,
ymajorgrids,
yminorgrids,
legend style={font=\footnotesize,at={(0.01,0.01)},nodes={scale=0.75, transform shape}, anchor=south west, legend cell align=left, align=left, draw=white!25!black, opacity=0.9}
]
\addplot [color=red, dashdotted, line width=1.0pt]
  table[row sep=crcr]{%
-33.0102999566398	16.4931959371976\\
-23.0102999566398	5.21560650569797\\
-13.0102999566398	1.64931959371976\\
-3.01029995663981	0.521560650569797\\
6.98970004336019	0.164931959371976\\
16.9897000433602	0.0521560650569798\\
26.9897000433602	0.0164931959371976\\
36.9897000433602	0.00521560650569797\\
46.9897000433602	0.00164931959371976\\
};
\addlegendentry{Non-coherent PEB for P1}

\addplot [color=black, dashdotted, line width=1.0pt]
  table[row sep=crcr]{%
-33.0102999566398	14.8438763434778\\
-23.0102999566398	4.69404585512817\\
-13.0102999566398	1.48438763434778\\
-3.01029995663981	0.469404585512817\\
6.98970004336019	0.148438763434778\\
16.9897000433602	0.0469404585512818\\
26.9897000433602	0.0148438763434778\\
36.9897000433602	0.00469404585512817\\
46.9897000433602	0.00148438763434778\\
};
\addlegendentry{Non-coherent PEB for P2}

\addplot [color=red, line width=1.0pt]
  table[row sep=crcr]{%
-33.0102999566398	0.0563721743987168\\
-23.0102999566398	0.0178264467756178\\
-13.0102999566398	0.00563721743987168\\
-3.01029995663981	0.00178264467756178\\
6.98970004336019	0.000563721743987168\\
16.9897000433602	0.000178264467756178\\
26.9897000433602	5.63721743987168e-05\\
36.9897000433602	1.78264467756178e-05\\
46.9897000433602	5.63721743987168e-06\\
};
\addlegendentry{Coherent PEB for P1}

\addplot [color=black, line width=1.0pt]
  table[row sep=crcr]{%
-33.0102999566398	0.018052043162884\\
-23.0102999566398	0.00570855728143835\\
-13.0102999566398	0.0018052043162884\\
-3.01029995663981	0.000570855728143835\\
6.98970004336019	0.00018052043162884\\
16.9897000433602	5.70855728143835e-05\\
26.9897000433602	1.8052043162884e-05\\
36.9897000433602	5.70855728143835e-06\\
46.9897000433602	1.8052043162884e-06\\
};
\addlegendentry{Coherent PEB for P2}

\addplot [color=red, line width=1.0pt, mark size=2.8pt, mark=diamond, mark options={solid, red}]
  table[row sep=crcr]{%
-33.0102999566398	2.54165635106258\\
-23.0102999566398	2.18132034284652\\
-13.0102999566398	1.85403291181665\\
-3.01029995663981	0.958134530562849\\
6.98970004336019	0.265610971793729\\
16.9897000433602	0.0422357428010014\\
26.9897000433602	5.63721743987168e-05\\
36.9897000433602	1.78264467756178e-05\\
46.9897000433602	5.63721743987168e-06\\
};
\addlegendentry{RMSE for P1}

\addplot [color=black, line width=1.0pt, mark size=2.8pt, mark=diamond, mark options={solid, black}]
  table[row sep=crcr]{%
-33.0102999566398	2.32682951277135\\
-23.0102999566398	1.94173865474599\\
-13.0102999566398	0.984813002955584\\
-3.01029995663981	0.288744153237508\\
6.98970004336019	0.0902329114494876\\
16.9897000433602	6.41244728143835e-05\\
26.9897000433602	2.08052043162884e-05\\
36.9897000433602	5.94527085572814e-06\\
46.9897000433602	1.46580520431629e-06\\
};
\addlegendentry{RMSE for P2}

\end{axis}
\end{tikzpicture}%
    \vspace{-3em}
        \caption{RMSE of the UE position estimate, and PEB for transmission protocols \textit{P1} and \textit{P2}.}
     	\label{fig:RMSE_PEB_vs_tx_power} 
\end{figure}
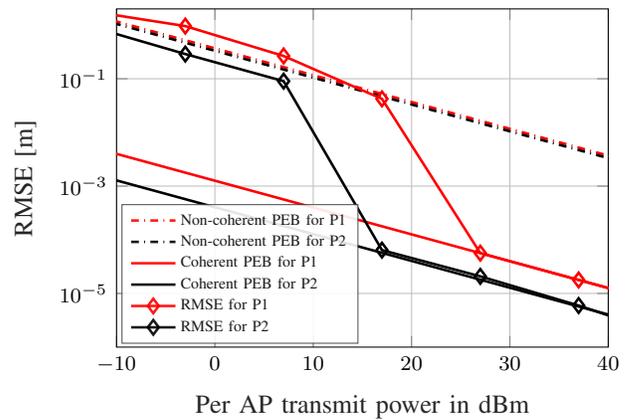

\begin{figure}
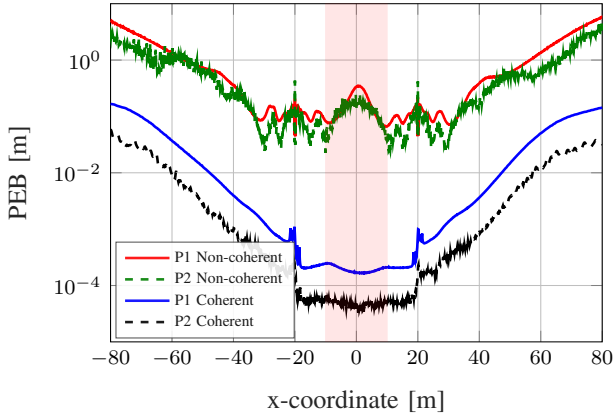

    \centering
    \include{TikzFigures/PEB_vs_x_axis_Coherent_Non_cohernet}
    \vspace{-3em}
        \caption{PEB values versus the x-coordinate of the UE position for transmission protocols \textit{P1} and \textit{P2} with coherent and non-coherent processing. The illumination region is shown as red shaded area. }
     	\label{fig:PEB_vs_x-axis}
\end{figure}

\subsection{PEB Analysis}
Fig. \ref{fig:PEB_vs_x-axis} shows the 1D PEB as a function of the $x$-coordinate of the UE position, for $y=1$ m, considering both coherent and non-coherent processing for protocols \textit{P1} and \textit{P2}. The illumination region, which is indicated by the vertical dotted lines in Fig. \ref{fig:PEB_vs_x-axis}, represents the portion of space where APs focus (\textit{P2}) or point (\textit{P1}) their beams. We observe that for coherent processing, the PEB values for both \textit{P1} and \textit{P2} reach their minima at the center of the illumination region. In contrast, for non-coherent processing, the PEB at the center is higher than at the edges. This occurs because the beams formed by APs in non-coherent operation are much wider than those in coherent operation, resulting in limited spatial diversity at the center (illuminated by the mainlobe, unlike the edges illuminated by the sidelobes), which degrades positioning performance \cite{diversity_localization_2020}. As expected, the PEB monotonically increases as the UE moves beyond the illumination region, since the available signal power decreases moving far away from the illumination region. Outside the illumination region, the received signal primarily originates from the side lobes of the beams, whose energy decreases with increasing distance.


\begin{figure}[t]
    \centering
%
%
\definecolor{mycolor1}{rgb}{0.00000,0.49804,0.00000}%
\begin{tikzpicture}[scale=1\columnwidth/9cm,font=\footnotesize]
\begin{axis}[%
width=0.75\columnwidth,
height=1.5in,
at={(0.755in,0.524in)},
scale only axis,
xmin=3,
xmax=65,
xlabel style={font=\color{white!15!black}},
xlabel={Number of illumination points $K$},
ymode=log,
ymin=1e-06,
ymax=100000000,
yminorticks=true,
ylabel style={font=\color{white!15!black}},
ylabel={PEB  [m]},
axis background/.style={fill=white},
xmajorgrids,
ymajorgrids,
yminorgrids,
legend style={font=\small,at={(0.55,0.6)},nodes={scale=0.75, transform shape}, anchor=south west, legend cell align=left, align=left, draw=white!25!black}
]
\addplot [color=red, line width=1.0pt]
  table[row sep=crcr]{%
4	0.5321\\
9	0.4123\\
16	0.33752\\
25	0.3173\\
36	0.3004\\
49	0.2821\\
64	0.283652\\
};
\addlegendentry{P1 Non-coherent}

\addplot [color=mycolor1, line width=1.pt, dashed]
  table[row sep=crcr]{%
4	0.0003\\
9	0.000276\\
16	0.000216\\
25	0.000202\\
36	0.000194\\
49	0.000176\\
64	0.000176522469\\
};
\addlegendentry{P1 Coherent}

\addplot [color=blue, line width=1.0pt]
  table[row sep=crcr]{%
4	4070000\\
9	439000\\
16	279\\
25	0.2924\\
36	0.1807\\
49	0.1745\\
64	0.1518\\
};
\addlegendentry{P2 Non-Coherent}

\addplot [color=black, line width=1.pt, dashed]
  table[row sep=crcr]{%
4	1600\\
9	819\\
16	0.00815\\
25	5.66e-05\\
36	4.8e-05\\
49	4.4e-05\\
64	4.4e-05\\
};
\addlegendentry{P2 Coherent}

\end{axis}
\end{tikzpicture}%
    \vspace{-3em}
        \caption{ PEB for different number of illumination points ($K$).}
     	\label{fig:PEB_vs_IP}
\end{figure}
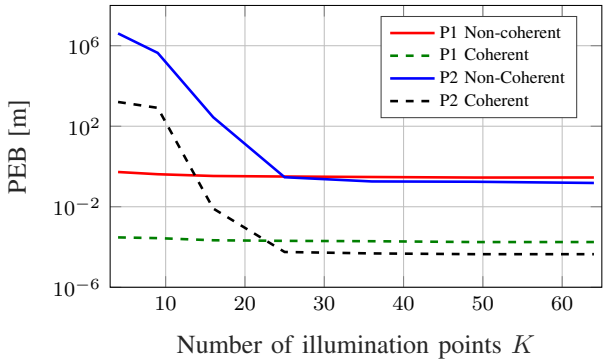


An important insight into the performance of the two protocols emerges from examining how the PEB varies with the number of illumination points, $K$. Fig. \ref{fig:PEB_vs_IP} illustrates the average PEB in the illumination region as a function of the number of illumination points. We observe that, for \textit{P1}, the PEB values remain nearly constant for any number of illumination points. Indeed, in \textit{P1}, the APs transmit sequentially, generating broad beams. This covers the entire illumination region, even when the number of illumination points is small. In contrast, for \textit{P2}, the PEB is significantly higher when the number of illumination points is low. In this case, when the UE is far away from an illumination point, it receives a signal that is due to the sidelobes of the \textit{coherent power pattern} generated by the AP network, that rapidly decrease around each illumination point. This is equivalent of having illumination gaps that, on average, increase the PEB. As the number of illumination points increases, these gaps progressively reduce and the PEB consequently decreases, stabilizing around a slightly smaller value compared to P1. The trend of the PEB with respect to the number of illumination points allows selecting the minimum value of $K$ for which protocol \textit{P2} is advantageous compared to \textit{P1}, yielding similar PEB performance at a lower latency.


\section{Conclusion}
This paper investigated joint localization and synchronization in the downlink of a distributed MIMO system with phase-coherent access points. We developed maximum likelihood estimators for two transmission protocols—sequential (\textit{P1}) and simultaneous (\textit{P2})—and analyzed their performance under coherent and non-coherent processing. Our results demonstrate that \textit{P2} achieves superior localization accuracy and lower latency. Furthermore, we  introduced a hybrid optimization approach to enhance estimation robustness. These findings highlight the potential of phase-coherent downlink D-MIMO for high-precision localization in future 6G networks.

\bibliographystyle{IEEEtran}
\bibliography{IEEEabrv,references}
\end{document}